\documentclass[twocolumn,showpacs,preprintnumbers,amsmath,amssymb]{revtex4}
\usepackage[dvips]{graphicx}
\usepackage{longtable}
\usepackage{color}

\def\ket#1{  \left\vert  #1   \right\rangle   }
\def\bra#1{  \left\langle  #1   \right\vert   }
\def\etal{\textit{et al.}}

\begin{document}

\title{Entanglement dynamics of three-qubit states in many-sided noisy channels}

\author{Michael Siomau}
 \email{siomau@physi.uni-heidelberg.de}
 \affiliation{Physikalisches
Institut, Heidelberg Universit\"{a}t, D-69120 Heidelberg, Germany}
 \affiliation{Department of Theoretical Physics, Belarussian State University, 220030 Minsk,
Belarus}

\begin{abstract}
We study entanglement dynamics of pure three-qubit
Greenberger-Horne-Zeilinger-type (GHZ-type) entangled states when
one, two or three qubits being subjected to general local noise.
Employing a lower bound for three-qubit concurrence as an
entanglement measure, we show that for some many-sided noisy
channels the entanglement dynamics can be completely described by
the evolution of the entangled states in single-sided channels.
\end{abstract}

\pacs{03.67.Mn, 03.65.Yz.}

\maketitle

\section{\label{sec:1} Introduction}

It is widely accepted nowadays that entangled states of
multiparticle systems are the most promising resource for quantum
information processing. At the same time, entanglement of complex
systems is known to be very fragile with regard to decoherence,
which may appear, for instance, due to a transmission of the whole
quantum system or some its subsystems through communication
channels. In practice, moreover, it is often required to distribute
parts of an entangled multipartite system between several remote
recipients. In this case, each subsystem is coupled locally with its
environment. Such a coupling of a quantum subsystem to some
environmental channel leads to decoherence of the multipartite
system and usually to some loss of entanglement. For successful
practical utilization of entangled states, it is of great importance
to quantify entanglement of the complex quantum systems and describe
quantitatively their entanglement dynamics under the action of local
(noisy) channels.

There are two main approaches in the literature to investigate
entanglement dynamics of multiparticle entangled states. First one
is to study state evolution of particular (usually maximally
entangled) states under the action of some chosen noisy channels and
deduce entanglement dynamics from the state evolution
\cite{Carvalho:04,Borras:09,Liu:09,Siomau1:10}. This approach,
however, is restricted by our choice of the entangled states and the
noise models. Recently, a completely different approach for the
description of entanglement dynamics, which is based on evolution
equation for entanglement, has been developed
\cite{Konrad:08,Li(d):09,Gour:10,Siomau2:10}. This latter approach
allows to obtain a direct relationship between the initial and the
final entanglement of the system in which just one of its subsystems
is subjected to an arbitrary noise. Unfortunately, the suggested
evolution equations can not be straightforwardly generalized to the
case when more than one subsystem undergo the action of local noisy
channels.

In this work, we investigate entanglement dynamics of initially pure
three-qubit states when one, two or three qubits are affected by
general local single-qubit noisy channels. At first, we shall
consider the entanglement dynamics of the maximally entangled
Greenberger-Horne-Zeilinger (GHZ) state, which can be written, in
the computational basis, as
\begin{equation}
  \label{GHZ}
\ket{\rm GHZ} = \frac{1}{\sqrt{2}} \left( \ket{000} + \ket{111}
\right) \, .
\end{equation}
Later, we shall generalize our discussion to all (GHZ-type) pure
three-qubit states which can be obtained from the state (\ref{GHZ})
by local unitary transformations. To describe the influence of the
local noisy channels on the entangled states we shall use quantum
operation formalism \cite{Nielsen:00}. Since there is no an
analytically computable measure of entanglement for multiqubit
states \cite{Horodecki:09}, a lower bound for multiqubit concurrence
\cite{Li(b):09} will be utilized to access the entanglement
dynamics. We shall show that for some noisy channels the complex
two- and three-sided entanglement dynamics, i.e. when two or three
qubits are affected by local noise, can be completely described by
the evolution of the entangled system in single-sided channels (when
just one qubit is subjected to local noise).

This work is organized as follows. In the next section, we present
the entanglement measure of use, the lower bound for multiqubit
concurrence, and recall its properties. In Sec. \ref{sec:3}, we step
by step analyze cases when one, two or three qubits of the
three-qubit system are subjected to local noisy channels. A summary
is drawn in Sec. \ref{sec:4}.

\section{\label{sec:2} The entanglement measure}

Concurrence, originally suggested by Wootters \cite{Wootters:98} to
describe entanglement of an arbitrary state of two qubits, has been
recognized as a very powerful measure of entanglement. Although
various extensions of the concurrence to the case of bipartite
states, if the dimensions of the associated Hilbert (sub-)spaces are
larger than two, has been suggested
\cite{Coffman:00,Rungta:01,Gour:05}, a full generalization of the
concurrence towards multiparticle states still remains challenging
\cite{Horodecki:09}. However, a formal extension of the concurrence
to multipartite case can be successfully approximated by an
analytically computable function, so-called lower bound, which never
exceeds such multipartite concurrence, but nonetheless is close
enough to its values. To date, several lower bounds for the
multipartite concurrence have been proposed
\cite{Carvalho:04,Horodecki:09,Li(b):09}. Let us recall and exploit
the lower bound for the multiqubit concurrence as suggested by  Li
\etal \, \cite{Li(b):09}.

The multiqubit concurrence for a pure three-qubit state $\ket{\psi}$
is given by
\begin{equation}
  \label{c-pure}
C_3(\ket{\psi}) = \sqrt{1 - \frac{1}{3} \sum_{i=1}^3 {\rm Tr}\,
\rho_i^2 } \, ,
\end{equation}
where the $\rho_i = {\rm Tr} \ket{\psi}\bra{\psi}$ denote the
reduced density matrix of the $i$-th qubit which is obtained by
tracing out the remaining two qubits. The concurrence for an
arbitrary mixed three-qubit state can be formally defined by means
of the so-called convex roof
\begin{equation}
  \label{c-mixed}
C_3(\rho) = {\rm min} \sum_i p_i \, C_3(\ket{\psi_i}) \, ,
\end{equation}
relying on the fact that any mixed state can be expressed as a
convex sum of some pure states $\{ \ket{\psi_i} \}$: $\rho = \sum_i
\, p_i \ket{\psi_i}\bra{\psi_i}$. However, the minimum in this
expression is to be found among all possible decompositions of
$\rho$ into pure states $\ket{\psi_i}$. No solution has been found
so far to optimize the concurrence (\ref{c-mixed}) analytically
\cite{Horodecki:09}.

A simple analytically computable lower bound $\tau_3 (\rho)$ for
three-qubit concurrence can be given in terms of the three bipartite
concurrences $C^{ab|c}$ ($a,b,c=1,2,3$ and $a\neq b\neq c\neq a$)
\cite{Li(b):09} as
\begin{equation}
  \label{low-bound-three}
\tau_3 (\rho) = \sqrt{\frac{1}{3}\, \sum_{k=1}^6 \, (C_k^{12|3})^2
+ (C_k^{13|2})^2 + (C_k^{23|1})^2 } \, .
\end{equation}
Here, each bipartite concurrence $C^{ab|c}$ corresponds to a
possible (bipartite) cut of the three-qubit system in which just one
of the qubits is discriminated from the other two qubits. For a
separation $ab|c$, the bipartite concurrence $C^{ab|c}$ is given by
a sum of six terms $C_k$ which are expressed as
\begin{equation}
 \label{concurence}
C_k^{ab|c} = {\rm max} \left( 0, \lambda_k^1 - \lambda_k^2 -
\lambda_k^3 - \lambda_k^4 \right) \, ,
\end{equation}
and where the $\lambda_k^m, \, m=1..4$ are the square roots of the
four nonvanishing eigenvalues of the matrix $\rho\,
\tilde{\rho}_k^{\: ab|c}$, if taken in decreasing order. These
matrices $\rho\: \tilde{\rho}_k^{\: ab|c}$ are formed by means of
the density matrix $\rho$ and its complex conjugate $\rho^*$, and
are further transformed by the operators $\{ S_k^{ab|c} = L^{ab|c}_k
\otimes L_0,\; k = 1,...,6 \}$ as: $\tilde{\rho}_k^{n} = S_k^{ab|c}
\rho^\ast S_k^{ab|c}$. In this notation, moreover, $L_0$ is the
(single) generator of the group SO(2) which is given by the second
Pauli matrix $\sigma_y = - i \, ( \ket{0}\bra{1} + \ket{1}\bra{0})$;
while the $\{ L^{ab|c}_k \}$ are the six generators of the group
SO$(4)$ that can be expressed explicitly by means of the totally
antisymmetric Levi-Cevita symbol in four dimensions as
$(L_{kl})_{mn} = - i \, \varepsilon_{klmn}; \; k,l,m,n =1..4$
\cite{Jones:98}.

Since the lower bound (\ref{low-bound-three}) is just an
approximation for the convex roof (\ref{c-mixed}), it is of great
importance to know how accurate this bound is with respect to the
convex roof for multiqubit concurrence. It has been checked
\cite{Siomau:11} by sampling $100$ randomly generated density
matrices $\rho$ that the lower bound $\tau_3 (\rho)$ coincides with
the numerically simulated convex roof $C_3(\rho)$ for all density
matrices with rank $r \leq 4$. This result has found its explanation
in the fact that the lower bound $\tau_3 (\rho)$, by its
construction \cite{Li(b):09}, relies only on four nonvanishing
eigenvalues of the matrix $\rho\, \tilde{\rho}_k^{\: ab|c}$, while
this matrix may have at most eight eigenvalues.

Taking into account the mentioned property of the lower bound
$\tau_3 (\rho)$, it is convenient to define a function of at most
four variables
\begin{eqnarray}
 \label{func}
f(w,x,y,z) & = &
 \nonumber
 \\[0.1cm]
& & \hspace*{-1.4cm} {\rm min} \left( 0, \; 2\,{\rm
max}\left(w,x,y,z\right) -w -x -y -z \right) \, .
\end{eqnarray}
It is easy to see that $C_k^{ab|c} = f(\lambda_k^1, \lambda_k^2,
\lambda_k^3, \lambda_k^4)$, where $\lambda_k^m, \, m=1..4$ are the
square roots of the eigenvalues of the matrix $\rho\,
\tilde{\rho}_k^{\: ab|c}$. Whenever the matrix $\rho\,
\tilde{\rho}_k^{\: ab|c}$ has less than four, e.g. two, eigenvalues,
we shall use the function of just two inputs $f(w, x) \equiv f(w, x,
0, 0)$.

\section{\label{sec:3} Entanglement dynamics in local noisy channels}

Quantum operation formalism is a very general and prominent tool to
describe how a quantum system has been influenced by its
environment. According to this formalism the final state of the
quantum system, that is coupled to some environmental channel, can
be obtained from its initial state with the help of (Kraus)
operators
\begin{equation}
 \label{sum-represent}
 \rho_{\rm fin} = \sum_i K_i \, \rho_{\rm ini} \, K_i^\dag \, ,
\end{equation}
and the condition $\sum_i K_i^\dag \, K_i = I$ is fulfilled. Note
that we consider only such system-environment interactions that can
be associated with completely positive {\it trace-preserving} maps
\cite{Nielsen:00}.

If the quantum system of interest consists of just single qubit
which is subjected to some environmental channel $A$, then an
arbitrary quantum operation can be expressed with the help of at
most four operators \cite{Nielsen:00}. Let us define the four
operators through the Pauli matrices as
\begin{eqnarray}
  \label{single}
 K_1(a_1) &=& \frac{a_1}{\sqrt{2}} \left( \begin{array}{cc} 1 & 0 \\ 0 & 1
\end{array} \right) , \;
 K_2(a_2) = \frac{a_2}{\sqrt{2}} \left( \begin{array}{cc} 0 & 1 \\1 &
0 \end{array} \right) \, , \nonumber
 \\[0.1cm]
 K_3(a_3) &=& \frac{a_3}{\sqrt{2}} \left( \begin{array}{cc} 0 & -i \\ i & 0
\end{array} \right) , \:
 K_4(a_4) = \frac{a_4}{\sqrt{2}} \left( \begin{array}{cc} 1 & 0 \\0 & -1
\end{array} \right) \,
\end{eqnarray}
where $a_i$ are real parameters and the condition $\sum_{i=1 }^4
a_i^2 = 1$ holds.

For three-qubit system, local interactions of the qubits with
channels $A,B$ and $C$ can be described by operators, which are
constructed as tensor products of the single-qubit operators
$K_i(a_i), K_i(b_i)$ and $K_i(c_i)$. Therefore, if the three qubits
are affected by local noise simultaneously, the final state of the
system (\ref{sum-represent}) can be obtained from its initial state
with the help of $64$ operators $K_i(a_i)\otimes K_j(b_j)\otimes
K_l(c_l)$ $(i,j,l=1...4)$. If just two qubits undergo the action of
local channels, the total number of Kraus operators is $16$.

\subsection{\label{sec:3.1} Single-sided channels}

Suppose, just one qubit of the three qubit system, which is
initially prepared in pure $GHZ$ state (\ref{GHZ}), is subjected to
a noisy channel $A$. The final state of the system is in general
mixed and can be expressed by a rank-4 density matrix $\rho$. This
matrix is obtained from Eq.~(\ref{sum-represent}) using the four
operators $1\otimes 1 \otimes K_i(a_i)$. Here we assumed, without
loss of generality, that the third qubit undergoes the action of the
channel. From the matrix $\rho$ we can compute the three bipartite
concurrences $C^{12|3}, C^{13|2}, C^{23|1}$ and the lower bound
(\ref{low-bound-three}). The concurrences are given by
\begin{eqnarray}
 \label{1-side-1}
C^{12|3} &=& f(a_1^2, a_2^2, a_3^2, a_4^2) \, ,
 \\[0.1cm]
\label{1-side-2}
 C^{13|2} &=& \sqrt{f^2(a_1^2, a_4^2) + f^2(a_2^2,
a_3^2)} \, ,
 \\[0.1cm]
\label{1-side-3}
 C^{23|1} &=& \sqrt{f^2(a_1^2, a_4^2) + f^2(a_2^2,
a_3^2)} \, ,
\end{eqnarray}
where we dropped the arguments of the concurrences, e.g. $C^{12|3}
\equiv  C^{12|3} (\left[1\otimes 1\otimes A\right]
\ket{GHZ}\bra{GHZ})$

Interestingly, if the channel $A$ is the bit-flip $(a_1\neq 0,
a_2\neq 0, a_3=a_4=0)$ or the bit-phase-flip $(a_1\neq 0, a_3\neq 0,
a_2=a_4=0)$ channel \cite{Nielsen:00}, the entanglement of the
three-qubit system, described by means of the lower bound $\tau_3
(\rho)$, never vanish.

The description of the entanglement dynamics of the $GHZ$ state
(\ref{GHZ}) in single-sided channels can be generalized to the case
of an arbitrary pure three-qubit state $\ket{\psi}$ which can be
obtained from the $GHZ$ state by local unitary transformations. Such
an extension is possible due to recently suggested evolution
equation for the lower bound $\tau_3$ \cite{Siomau2:10}, which
manifests that the entanglement dynamics of an arbitrary pure state
$\ket{\psi}$ of a three-qubit system, when one of its qubits
undergoes the action of an arbitrary noisy channel $A$, is
subordinated to the dynamics of the maximally entangled state, i.e.
\begin{eqnarray}
 \label{tau}
\tau_3 [(1 \otimes 1 \otimes A) \ket{\psi}\bra{\psi}] &=&
 \nonumber
 \\[0.1cm]
& & \hspace*{-2.8cm} \tau_3 [(1 \otimes 1 \otimes A)
\ket{GHZ}\bra{GHZ}] \: \tau_3 [\ket{\psi}] \, .
\end{eqnarray}
Conclusively, the entanglement dynamics of an arbitrary pure
three-qubit state, which is locally equivalent to the GHZ state and
is subjected to an arbitrary single-sided channel $A$, is given by
the lower bound (\ref{low-bound-three}), which is defined through
Eqs.~(\ref{1-side-1})-(\ref{1-side-3}), and the evolution equation
(\ref{tau}). Moreover, based on the numerical simulations in
\cite{Siomau:11}, we argue that the description of this particular
case of the entanglement dynamics with the lower bound $\tau_3$ is
as general as with the convex roof (\ref{c-mixed}) for multiqubit
concurrence.

\subsection{\label{sec:3.2} Two-sided channels}

If the three-qubit system is prepared into the $GHZ$ state
(\ref{GHZ}) and just two (let say, the second and the third) qubits
are affected by local channels $A$ and $B$ simultaneously, the final
state of the system is given by a rank-8 density matrix $\rho$. In
this case, the three bipartite concurrences can be expressed as
\begin{widetext}
\begin{small}
\begin{eqnarray}
 \label{2-side-1}
C^{12|3} = \sqrt{f^2(a_2^2b_3^2+a_3^2b_2^2, a_2^2b_2^2+a_3^2b_3^2,
a_3^2b_1^2+a_2^2b_4^2, a_2^2b_1^2+a_3^2b_4^2)
+f^2(a_4^2b_2^2+a_1^2b_3^2, a_1^2b_2^2+a_4^2b_3^2,
a_4^2b_1^2+a_1^2b_4^2, a_1^2b_1^2+a_4^2b_4^2 )} \, ,
 \\[0.1cm]  \label{2-side-2}
C^{13|2} = \sqrt{f^2(a_4^2b_2^2 +a_1^2b_3^2, a_3^2b_2^2 +
a_2^2b_3^2, a_2^2b_2^2 +a_3^2b_3^2, a_1^2b_2^2 +a_4^2b_3^2 )
+f^2(a_4^2b_1^2 +a_1^2b_4^2, a_3^2b_1^2 +a_2^2b_4^2, a_2^2b_1^2
+a_3^2b_4^2, a_1^2b_1^2 +a_4^2b_4^2)} \, ,
 \\[0.1cm]  \label{2-side-3}
C^{23|1} = \sqrt{f^2(a_4^2b_2^2 +a_1^2b_3^2, a_1^2b_2^2 +a_4^2b_3^2,
a_3^2b_1^2 +a_2^2b_4^2, a_2^2b_1^2 +a_3^2b_4^2) +f^2(a_3^3b_2^2
+a_2^2b_3^2, a_2^2b_2^2+a_3^2b_3^2, a_4^2b_1^2 +a_1^2b_4^2,
a_1^2b_1^2 +a_4^2b_4^2)} \, .
\end{eqnarray}
\end{small}
\end{widetext}
Although no general conclusion about the entanglement dynamics of
the three-qubit system can be made from the structure of these
bipartite concurrences, there are some interesting particular cases.
If each of the channels $A$ and $B$ is just the bit-flip or the
bit-phase-flip channel, the squared lower bound $\tau_3^2$ is given
by
\begin{equation}
  \label{low-bound-2-sided}
\frac{1}{3} \left[ f^2(a_1^2, a_i^2) + f^2(b_1^2, b_j^2) +
f^2(a_1^2, a_i^2)f^2(b_1^2, b_j^2) \right] \, ,
\end{equation}
where $i,j = 2,3$. Symbolically Eq.~(\ref{low-bound-2-sided}) can be
written as
\begin{equation}
  \label{gen}
\tau_3^2 (\left[ 1\otimes A\otimes B\right]\rho) = \frac{1}{3} \,
\left( \mathbb{A}^2 + \mathbb{B}^2 + \mathbb{A}^2
\mathbb{B}^2\right) \, ,
\end{equation}
where $\mathbb{A} \equiv C^{12|3}(\left[1\otimes 1\otimes
A\right]\rho), \, \mathbb{B} \equiv C^{12|3}(\left[1\otimes 1\otimes
B\right]\rho)$ and $\rho=\ket{GHZ}\bra{GHZ}$. Other words,
entanglement dynamics of the $GHZ$ state in two-sided noisy
channels, when the channels $A$ and $B$ are the bit-flip or the
bit-phase-flip channels, is completely defined by the dynamics of
the state in the single-sided channels. Although this result holds
only for mentioned channels, it can be extended to the case of an
arbitrary pure three-qubit state $\rho = \ket{\psi}\bra{\psi}$ which
is locally equivalent to the $GHZ$ state. This extension is based on
the evolution equation for bipartite concurrence \cite{Li(d):09}
which has very similar sense and structure to the evolution equation
(\ref{tau}) for the lower bound. For the bipartite concurrence
$C^{12|3}$, for example, the evolution equation is given by
\begin{eqnarray}
 \label{tau}
C^{12|3} [(1 \otimes 1 \otimes A) \ket{\psi}\bra{\psi}] &=&
 \nonumber
 \\[0.1cm]
& & \hspace*{-4cm} C^{12|3} [(1 \otimes 1 \otimes A)
\ket{GHZ}\bra{GHZ}] \: C^{12|3} [\ket{\psi}] \, .
\end{eqnarray}

It is important to note that Eqs.~(\ref{2-side-1})-(\ref{2-side-3})
are computed for a rank-8 density matrix and, therefore, the lower
bound $\tau_3$ obtained with the help of these equations may differ
from the convex roof (\ref{c-mixed}) for the concurrence
significantly. However, due to the assumption that the channels $A$
and $B$ are the bit-flip or the bit-phase-flip channels, the state
$\left[1\otimes A\otimes B\right]\ket{GHZ}\bra{GHZ}$ has rank four.
Therefore, it can be argued that Eq.~(\ref{gen}), formulated for the
lower bound $\tau_3$, remains valid if the convex roof
(\ref{c-mixed}) is substituted in its left hand side (lhs).

\subsection{\label{sec:3.3} Three-sided channels}

In the previous subsection we have analyzed the case when just two
qubits of the three-qubit system are subjected to local noisy
channels. Similar analysis can be made when all three qubits are
affected by local channels $A, B$ and $C$ simultaneously. If the
system is initially prepared in the $GHZ$ state (\ref{GHZ}), the
final state density matrix, derived from Eq.~(\ref{sum-represent}),
has rank eight. The analytically computed concurrences $C^{12|3},
C^{13|2}$ and $C^{23|1}$ in this case have very complex structure
and, therefore, we do not display them here. There are significant
simplifications in the structure of the concurrences and the lower
bound if all $A, B$ and $C$ are bit-flip channels or if two of these
channels are bit-phase-flip and the remaining one is a bit-flip. In
these two cases, the squared lower bound $\tau_3^2$ can be written
as
\begin{equation}
  \label{gen2}
\tau_3^2 (\left[ A\otimes B\otimes C\right]\rho) = \frac{1}{3} \,
\left( \mathbb{A}^2 \mathbb{B}^2 + \mathbb{B}^2 \mathbb{C}^2 +
\mathbb{A}^2 \mathbb{C}^2\right) \, ,
\end{equation}
where $\mathbb{A} \equiv C^{12|3}(\left[1\otimes 1\otimes
A\right]\rho)$, $\mathbb{B} \equiv C^{12|3}(\left[1\otimes 1\otimes
B\right]\rho)$, $\mathbb{C} \equiv C^{12|3}(\left[1\otimes 1\otimes
C\right]\rho)$ and $\rho=\ket{GHZ}\bra{GHZ}$. The entanglement
dynamics of the $GHZ$ state in the three-sided noisy channels is
given just by the dynamics of the state in the single-sided
channels. As in the previous section, this result can be generalized
to the case of an arbitrary three-qubit state (which is locally
equivalent to the $GHZ$ state) due to the evolution equation for
bipartite concurrence \cite{Li(d):09}. Moreover, the fact that the
state $\left[A\otimes B\otimes C\right] \ket{GHZ}\bra{GHZ}$ has rank
four, for mentioned channels, suggests to substitute the convex roof
(\ref{c-mixed}) in the lhs of Eq.~(\ref{gen2}) without loss of any
information about the entanglement dynamics.

\section{\label{sec:4} Summary}

Our analysis in the previous section suggests that the entanglement
dynamics of a pure three-qubit GHZ-type entangled state $\ket{\psi}$
in some local many-sided noisy channels can be completely described
by evolution of the entangled state in single-sided channels. More
specifically, if just two qubits are subjected to local channels $A$
and $B$, so that the channel $\left[ 1\otimes A\otimes B\right]$ is
a combination of just bit-flip and/or bit-phase-flip channels, the
lower bound for three-qubit concurrence $\tau_3 (\left[ 1\otimes
A\otimes B\right]\ket{\psi}\bra{\psi})$ can be expressed without
loss of generality through the bipartite concurrences $
C^{12|3}(\left[1\otimes 1\otimes A\right]\ket{\psi}\bra{\psi})$ and
$C^{12|3}(\left[1\otimes 1\otimes B\right]\ket{\psi}\bra{\psi})$ by
Eq.~(\ref{gen}). If all three qubits undergo the action of local
noisy channels $A, B$ and $C$, which are neither bit-flip channels
or two of these channels are the bit-phase-flip and the remaining
one is the bit-flip, the lower bound $\tau_3 (\left[ A\otimes
B\otimes C\right]\ket{\psi}\bra{\psi})$ is again given just by the
bipartite concurrences as it is displayed in Eq.~(\ref{gen2}).

It is important to note that there are only two nonvanishing
eigenvalues in the expression (\ref{1-side-1}) for the bipartite
concurrence $C^{12|3}(\left[1\otimes 1\otimes
A\right]\ket{\psi}\bra{\psi})$, if the channel $A$ is the bit-flip
or the bit-phase-flip channel. Such a bipartite concurrence can be
directly measured \cite{Guehne:09}. Eqs.~(\ref{gen}) and
(\ref{gen2}) allow one to access the complex many-sided entanglement
dynamics just by measuring the bipartite concurrence in the simplest
case, when just one qubit is affected by a local noisy channel.

\end{document}